\documentstyle[epsfig,mnras_cite,times]{mn}
\newcommand{\citep}{\cite}
\title{The 2dF Galaxy Redshift Survey: The bias of galaxies
and the density of the Universe}
\author[Licia Verde et al.]{
\parbox[t]{\textwidth}{
Licia Verde$^{1,2}$,
Alan F. Heavens$^{3}$,
Will J. Percival$^{3}$,
Sabino Matarrese$^{4}$,
Carlton M.\ Baugh$^5$,
Joss Bland-Hawthorn$^6$,
Terry Bridges$^6$,
Russell Cannon$^6$,
Shaun Cole$^5$,
Matthew Colless$^7$,
Chris Collins$^8$,
Warrick Couch$^{9}$,
Gavin Dalton$^{10}$,
Roberto De Propris$^{9}$,
Simon P.\ Driver$^{11}$,
George Efstathiou$^{12}$,
Richard S.\ Ellis$^{13}$,
Carlos S.\ Frenk$^5$,
Karl Glazebrook$^{14}$,
Carole Jackson$^7$,
Ofer Lahav$^{12}$,
Ian Lewis$^{10}$,
Stuart Lumsden$^{15}$,
Steve Maddox$^{16}$,
Darren Madgwick$^{12}$,
Peder Norberg$^5$,
John A.\ Peacock$^{3}$,
Bruce A.\ Peterson$^7$,
Will Sutherland$^{3}$,
Keith Taylor$^6$}
\vspace*{6pt} \\
$^{1}$ Department of Physics and Astronomy, Rutgers University,
       136 Frelinghuysen Road, Piscataway, NJ 08854--8019 USA.\\
$^{2}$ Princeton University Observatory, Princeton, NJ 08544, USA.\\
$^{3}$ Institute for Astronomy, University of Edinburgh, Royal Observatory,
       Blackford Hill, Edinburgh EH9 3HJ, UK \\
$^{4}$ Dipartimento di Fisica, `G. Galilei', Universita' di
       Padova, and INFN, Sezione di Padova, via Marzolo 8, I-35131 Padova, Italy\\
$^5$Department of Physics, University of Durham, South Road,
    Durham DH1 3LE, UK \\
$^6$Anglo-Australian Observatory, P.O.\ Box 296, Epping, NSW 2121,
    Australia\\
$^7$Research School of Astronomy \& Astrophysics, The Australian
    National University, Weston Creek, ACT 2611, Australia \\
$^8$Astrophysics Research Institute, Liverpool John Moores University,
    Twelve Quays House, Birkenhead, L14 1LD, UK \\
$^{9}$Department of Astrophysics, University of New South Wales, Sydney,
    NSW 2052, Australia \\
$^{10}$Department of Physics, University of Oxford, Keble Road,
    Oxford OX1 3RH, UK \\
$^{11}$School of Physics and Astronomy, University of St Andrews,
    North Haugh, St Andrews, Fife, KY6 9SS, UK \\
$^{12}$Institute of Astronomy, University of Cambridge, Madingley Road,
    Cambridge CB3 0HA, UK \\
$^{13}$Department of Astronomy, California Institute of Technology,
    Pasadena, CA 91125, USA \\
$^{14}$Department of Physics \& Astronomy, Johns Hopkins University,
       Baltimore, MD 21218-2686, USA \\
$^{15}$Department of Physics, University of Leeds, Woodhouse Lane,
       Leeds, LS2 9JT, UK \\
$^{16}$School of Physics \& Astronomy, University of Nottingham,
       Nottingham NG7 2RD, UK \\
}

\newcommand{\be}{\begin{equation}}
\newcommand{\ee}{\end{equation}}

\newcommand{\ba}{\begin{eqnarray}}
\newcommand{\ea}{\end{eqnarray}}

\newcommand{\bk}{{\bf k}}

\newcommand{\br}{{\bf r}}

\newcommand{\bx}{{\bf x}}

\def\gs{\mathrel{\raise1.16pt\hbox{$>$}\kern-7.0pt 
\lower3.06pt\hbox{{$\scriptstyle \sim$}}}}         
\def\ls{\mathrel{\raise1.16pt\hbox{$<$}\kern-7.0pt 
\lower3.06pt\hbox{{$\scriptstyle \sim$}}}}         

\begin{document}

\maketitle

\begin{abstract}
We compute the bispectrum of the 2dF Galaxy Redshift Survey (2dFGRS)
and use it to measure the bias parameter of the galaxies.  This
parameter quantifies the strength of clustering of the galaxies
relative to the mass in the Universe. By analysing 80 million triangle
configurations in the wavenumber range $0.1 < k < 0.5\,h\,$Mpc$^{-1}$
(i.e. on scales roughly  between 5 and  30 $h^{-1}\,$ Mpc) we find that the
linear bias parameter is consistent with unity: $b_1=1.04\pm 0.11$,
and the quadratic (nonlinear) bias is consistent with zero:
$b_2=-0.054\pm 0.08$. Thus, at least on large scales, {\it
optically-selected galaxies do indeed trace the underlying mass
distribution}.  The bias parameter can be combined with the 2dFGRS
measurement of the redshift distortion parameter $\beta \simeq
\Omega_m^{0.6}/b_1$, to yield $\Omega_m = 0.27\pm 0.06$ for the matter
density of the Universe, a result which is determined entirely from
this survey, independently of other datasets. Our measurement of the matter
density of the Universe should be interpreted as $\Omega_m$ at the effective
redshift of the survey ($z=0.17$).
\end{abstract}


\section{Introduction}
Clustering of mass in the Universe is believed to be a result of
amplification by gravitational instability of small perturbations
generated in the early Universe.  Comparison with theoretical
predictions offers the chance to test models of the generation of the
perturbations, as well as putting important constraints on
cosmological parameters, which control the growth rate of the
perturbations.  A fundamental limitation on such a comparison has been
that theoretical models predict the clustering properties of the mass
in the Universe, and yet we have few direct measures of mass
observationally.  More readily observable is the distribution of
luminous objects such as galaxies, so to compare with theory one has
to determine, or assume, the relationship between the clustering of
mass and the clustering of galaxies. In general one will expect these
to differ, because the efficiency of galaxy formation may depend in
some nontrivial way on the underlying mass distribution.  The idea
that structures may be `biased' tracers of the mass distribution goes
back to \scite{Kai84}, who explained the high clustering strength of
Abell clusters as due to their forming in high-density regions of the
Universe.  In addition, observations indicating that different types
of galaxy cluster differently (e.g.
\pcite{Dress80,PG84,WTD88,Ham88,LNP90,LahavSaslaw92}) show that they
cannot all be unbiased tracers of the mass.  Bias became an attractive
way to reconcile the low velocities of galaxies with the high-density
Einstein-de Sitter model favoured in the 1980s (e.g. \pcite{DEFW85}),
but after the Cosmic Background Explorer (COBE) determined the
amplitude of primordial fluctuations on large scales \cite{COBE}, the
`standard' biased Cold Dark Matter (CDM) model became less popular.
With the advent of more detailed datasets in the cosmic microwave
background (CMB) and large-scale structure, it is possible to
investigate and constrain a wider range of galaxy formation models,
and an unknown bias relation adds uncertainty to the process.

Since the efficiency of galaxy formation is not well understood
theoretically, it makes sense to try to measure it empirically from
observations.  When the perturbations are small (or on large, linear
scales), it is difficult to do this: there is a degeneracy between the
unknown amplitude of the matter power spectrum $P(k)$ and the degree
of bias, $b$, defined such that the galaxy power spectrum is $P_g(k)
\equiv b^2 P(k)$.  In principle, $b$ may be a function of scale,
through the wavenumber $k$.  At later times (or on smaller scales),
however, the degeneracy is lifted by nonlinear effects.  One feature
of nonlinear gravitational evolution is that the overdensity field
$\delta(\bx) \equiv (\rho(\bx)-\overline{\rho})/\overline{\rho}\ $
becomes progressively more skewed towards high density.  In principle
skewness could also arise from non-Gaussian initial conditions; in
practice this can be neglected \cite{VWHK}, since CMB fluctuations are
consistent with Gaussian initial conditions
\cite{Komatsuetal01,Santosetal001}.  One can thus hope to exploit the
gravitational skewness, but skewness could equally well arise from
biasing, e.g. from a galaxy formation efficiency that increased at
dense points in the mass field.  It is nevertheless possible to
distinguish these two effects by considering the {\it shapes} of
isodensity regions.  If the field is unbiased, then the shapes of
isodensity contours become flattened, as gravitational instability
accelerates collapse along the short axis of structures, leading to
sheet-like and filamentary structures (e.g. \pcite{Zel70}).  If the
galaxy field is highly biased with the same power spectrum, however,
the underlying mass field is of low amplitude, and thus expected to be
close to the initial field, which is assumed to be Gaussian.  These
fields do not have highly-flattened isodensity contours, as bias does
not flatten the contours; for example Eulerian bias preserves the
contour shape.  Thus there is a difference which could be detected,
for example, by studying the three-point correlation function.  In
this paper, we exploit this effect in Fourier space rather than real
space, by analysing the bispectrum: $\langle
\delta_{\bk1}\delta_{\bk2} \delta_{\bk3} \rangle$, where $\delta_\bk$
is the Fourier transform of the galaxy overdensity field.  The theory
for the bispectrum is set out in \scite{Fry94}, \scite{HBCJ95},
\scite{MVH97}, \scite{VHMM98}, \scite{SCFFHM98}, \scite{SCF99} and
\scite{Scocc00}.

The galaxy survey we use is the Anglo-Australian Telescope 2
degree field Galaxy Redshift Survey \cite{Colless01}, as compiled
in February 2001.  It was created with the 2dF multi-fibre
spectrograph on the Anglo-Australian Telescope
\cite{Lewisetal2001}, and currently consists of over 200,000
galaxies with redshifts up to about $z=0.3$, broadly in two
regions centred near the south and north galactic poles.  See {\tt
http://www.mso.anu.edu.au/2dFGRS/} for further details. It is the
first survey which is large enough to put tight constraints on
the bias parameter, as previous surveys are too shallow or too
sparse.  In this paper, we use 127,000 galaxies from the February
2001 compilation of the catalogue, truncated at $0.03<z<0.25$.

The outline of the paper is as follows: in Section 2 we review
the theory of growth of the bispectrum through gravitational
instability, and discuss briefly the effects of redshift-space
distortions; in Section 3 we illustrate our method of measuring the
bias parameter. This method uses a new estimator of the bias
parameter, which allows us to analyse many millions
of $\bk$-vector triplets, thus drastically improving the signal to noise.
In Section 4 we test the performance of the method. Finally in
Section 5 we present our results and in Section 6 we discuss the
implications of these results. An Appendix presents and describes
in detail our new estimator of the bias parameter.

\section{The bispectrum in real and redshift space}

The statistic we use to measure the bias of the galaxies is the
bispectrum $B$, which is related to the three-point correlation
function in Fourier space.  For the mass, this is defined by
\begin{equation}
\langle \delta_{\bk_1}\delta_{\bk_2}\delta_{\bk_3}\rangle \equiv
(2\pi)^3 B(\bk_1,\bk_2,\bk_3) \delta^{\rm D}(\bk_1+\bk_2+\bk_3)
\label{bispmass}
\end{equation}
where $\delta_{\bk} \equiv \int d^3\bx\,
\delta(\bx)\exp(-i\bk\cdot\bx)$ is the Fourier transform of the
mass overdensity $\delta(\bx) \equiv \rho(\bx)/\overline{\rho}-1$ and 
$\delta^{\rm D}$ is the Dirac delta function, which shows that
the bispectrum can be non-zero only if the $\bk$-vectors close to
form a triangle.

The power spectrum $P$ is similarly defined by
\begin{equation}
\langle\delta_{\bk}\delta_{\bk'}\rangle \equiv (2\pi)^3 P(k) \delta^{\rm
D}(\bk+\bk').
\label{power}
\end{equation}
Analogous relations hold for the power spectrum and bispectrum of
the galaxy distribution. We assume that the mass overdensity is a
Gaussian random field initially, as closely predicted by
inflationary early universe models.  Thus, at asymptotically
early times the bispectrum is zero by symmetry. As gravitational
instability develops, the field becomes asymmetric, because nonlinear
effects skew the density field to high densities. In this way, a
non-zero bispectrum develops.  In the mildly nonlinear regime, we
use second-order perturbation theory to compute the expected
bispectrum.  To second order (in the overdensity $\delta$) the
Fourier coefficients develop a nonlinear component which is
proportional to $\delta^2$, so the leading order term in the
bispectrum grows like $\delta^4$.  Since the 2dFGRS is not
a survey of mass density, to interpret the bispectrum
measured from the survey we must make some assumption about the
distribution of mass relative to the distribution of galaxies. To
date, this uncertainty in the relationship between the mass and
the galaxy distribution has placed a limitation on the usefulness
of galaxy catalogues as a probe of cosmology.  We make the
assumption that the galaxy overdensity field $\delta_g$ is
related to the underlying mass overdensity by some deterministic
function, which we expand in a Taylor series as (cf.
\pcite{FG93})
\begin{equation}
\delta_g = \sum_{i=0}^\infty {b_i \delta^i\over i!}.
\label{eq:bias}
\end{equation}
We must keep terms up to $i=2$, since these enter in the
bispectrum at the same level as second-order perturbation theory
growth terms, and we ignore higher-order terms.  In order for
$\delta_g$ to have zero mean, there is a (calculable) $b_0$ term,
but we ignore it as it contributes only to $\bk = {\bf 0}$; $b_1$
is the linear bias parameter, and $b_2$ is the quadratic bias
parameter.  A non-zero $b_2$ would indicate nonlinear biasing of
galaxies with respect to mass. Both of these parameters are
estimated in this paper from the 2dFGRS.

In real space, the two effects of nonlinear growth and nonlinear
bias contribute terms to a non-zero bispectrum:
\begin{equation}
B(\bk_1,\bk_2,\bk_3) = P_g(k_1)P_g(k_2) \left[
{J(\bk_1,\bk_2)\over b_1} + {b_2\over b_1^2}\right] + {\rm cyc.}
\label{eq:bispectrum1}
\end{equation}
where there are two
additional cyclic terms (2,3) and (3,1).  Details of the theory
leading to (\ref{eq:bispectrum1}) may be found in e.g.
\scite{MVH97}. We assume here that the galaxy power spectrum is
$P_g(k) = b_1^2 P(k)$ (see \pcite{HMV98} for discussion of this
point); $J$ is a function that depends on the shape of the
triangle in $\bk-$space, but only very weakly on cosmology (e.g.
\pcite{BJCP92,BCHJ95,CLMM95}).  Note that we assume a
deterministic bias; other authors (e.g. \pcite{SW98};
\pcite{DekelLahav99}; \pcite{TKS99}) have investigated stochastic
bias, where there is a random component to the relationship
between $\delta$ and $\delta_g$. In the case of stochastic bias
the bispectrum has still the form of equation (3) but the function
$J$ is modified into $J'$ in such a way that when the correlation
coefficient of the stochastic bias $r$ goes to unity (i.e. bias
becomes deterministic), $J'\rightarrow J$. Theoretical
considerations suggest that $r \sim 1$ on scales of interest
(e.g. Tegmark \& Peebles 1998; Blanton et al. 2000; Seljak 2000).

In redshift space, both the power spectrum and the bispectrum are
modified by redshift-space distortions, arising because the
distance estimator (the redshift) is perturbed by peculiar
velocities.  These distortions are radial in nature, and can be
analysed, at some expense in complexity, in radial and angular
basis functions (e.g. \pcite{FSL94,HT95,BHT95,Ham98,Tadros99}).  Here,
we adopt the distant-observer approximation \cite{Kai87}, and
assume that nonlinear effects can be modelled by an incoherent
small-scale velocity field, characterised by the pairwise
velocity dispersion $\sigma_p$.  The large-scale in-fall leads to
distortions which depend on the redshift distortion parameter
$\beta = \Omega_m^{0.6}/b_1$, where $\Omega_m$ is the matter
density parameter.  Assuming in addition an exponential
distribution for the pairwise velocity the combined effect gives
the power spectrum in redshift space (denoted by subscript $s$)
\begin{equation}
P_s(\bk) = {P(k)(1+\beta\mu^2)^2\over 1+k^2\mu^2\sigma_p^2/2}
\label{Eq:redshiftpower}
\end{equation}
(e.g. \pcite{BPH96,HattonCole98}), where $\mu$ is the cosine of
the wavevector to the line of sight, which is a fixed direction
in the distant observer approximation. Other modifications have
been suggested, such as a Gaussian, or exponential one-particle
velocity dispersion, which yield different functional forms for
the redshift distortion. Note that $\sigma_p$ is usually written,
as here, implicitly divided by the Hubble constant. The bispectrum
is modified similarly, and again various modifications have been
proposed. We use the form
\begin{eqnarray}
B_s(\bk_1,\bk_2,\bk_3) & = & (B_{12}+B_{23}+B_{31})\times
\\\nonumber
& \!\!\! & \!\!\!\!\!\!\!\!\!\!\!\!\!\!\!\!\!\!\!\!\!\!\!\!\!\!\!\!\!
\!\!\!\!\!\!\!\!\!\!\!\!\!\!\!\!\!\!\!\!\!\!\!
\left[\!\left(1+{\alpha_V^2 k_1^2\mu_1^2\sigma_p^2\over
2}\right)\!\left(1+{\alpha_V^2 k_2^2\mu_2^2\sigma_p^2\over
2}\right)\!\left(1+{\alpha_V^2 k_3^2\mu_3^2\sigma_p^2\over
2}\right)\!\right]^{-1/2}
\label{BsVHMM}
\end{eqnarray}
where
\begin{eqnarray}
B_{12} & \equiv &
(1+\beta\mu_1^2)(1+\beta\mu_2^2)\times\\\nonumber
 & & \left[{{\rm Ker}(\bk_1,\bk_2)\over b_1} + {b_2\over b_1^2}
\right]P_{g}(\bk_1)P_{g}(\bk_2)
\label{B12}
\end{eqnarray}
and the kernel function ${\rm Ker}$ is $J$ modified for redshift
space (see \pcite{VHMM98} equation (13) for the formula);
$\alpha_V$ is an adjustable parameter which is shape-dependent,
and must be calibrated from simulations. \scite{SCF99} propose an
alternative\footnote{Note that in equation 38 of \scite{SCF99} there is
an extra power of 2 outside the square brackets, which we omit in
(\ref{Roman}).} for the denominator of (6), namely
\begin{equation}
\left(1+\alpha_S^2[k_1^2
\mu_1^2+k_2^2\mu_2^2+k_3^2\mu_3^2]\sigma_v^2/2\right)^2
\label{Roman}
\end{equation}
where the one-particle dispersion $\sigma_v=\sigma_p/\sqrt{2}$ if the
small-scale velocities are incoherent, and again the parameter
$\alpha_S$ needs to be calibrated for different triangle shapes and
cosmologies.  We find that the formula (6) recovers the true bias
parameter in an ensemble of simulated biased 2dFGRS catalogues with
smaller scatter than (\ref{Roman}).  We see from (6) and (7) how the
bispectrum can allow us to measure the bias parameters.  The left-hand
side is potentially observable, and we can hope to constrain $b_1$ and
$b_2$ by considering triangles of different shape (and hence different
${\rm Ker}$).

\section{Method}

The previous Section shows the theoretical model for the bispectrum,
and its dependence on the two parameters $b_1$ and $b_2$; note that,
apart from these two parameters, the bispectrum depends on observable
quantities such as $\beta$, $\sigma_p$, and $P_g$.  The real-space
galaxy power spectrum is obtained from the angle-averaged
redshift-space power (cf. \pcite{VHMM98}):
\begin{eqnarray}
P(k)& =&  P_s(k)\left[4\frac{(\sigma_p^2 k^2-\beta)\beta}{\sigma_p^4k^4}\right.+\\
\nonumber
 & & \left. \frac{2\beta^2}{3
\sigma_p^2k^2}+\frac{\sqrt{2}(k^2\sigma_p^2-2\beta)^2}{k^4\sigma_p^5}{\rm
 tan^{-1}}\left(\frac{k \sigma_p}{\sqrt{2}}\right) \right]^{-1}
\label{Eq:invertpower}
\end{eqnarray}
The real-space power spectrum obtained this way agrees well with
the APM power estimated by \scite{BE94a} and \scite{EM01}. The
input catalogue is based on a revised and extended version of the
APM galaxy catalogue \cite{Maddox90}.
\begin{figure*}
\begin{center}
\setlength{\unitlength}{1mm}
\begin{picture}(90,55)
\includegraphics{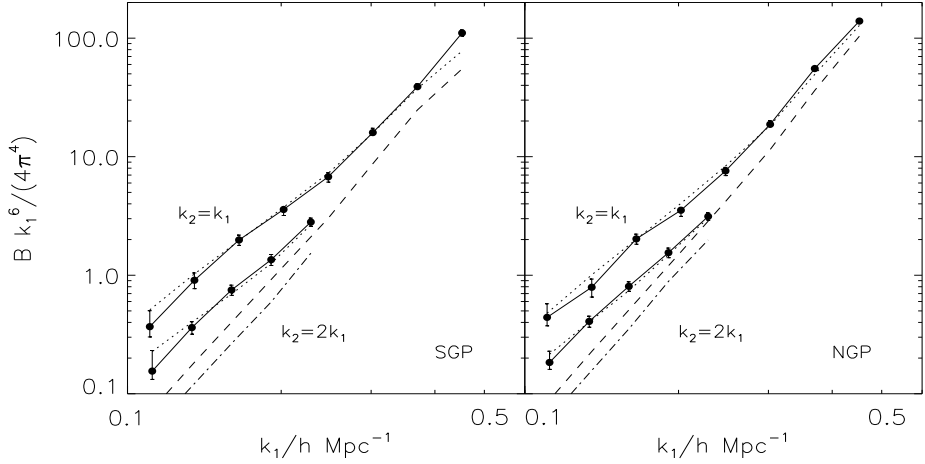}
\end{picture}
\end{center}
\caption{Measured (redshift-space) 2dFGRS dimensionless bispectrum
from the SGP and NGP for the two chosen configurations.  The
dotted line shows the perturbation theory prediction for $b_1=1$,
$b_2=0$ while the dashed and dot-dashed lines show the shot noise
contributions. The error bars are obtained via Monte Carlo
simulation of 16 mock 2dFGRS catalogues (see text for details).}
\label{fig:bispectrum}
\end{figure*}

The bispectrum and power spectrum data come from transforming the
galaxy distribution as follows.  The galaxies are weighted with
the optimum weight for measuring the power spectrum \cite{FKP94},
which also minimises the variance of higher-order correlation
functions \cite{Scocc00}.  The weight is $w(\br)=1/[1+P_0
\overline n(\br)]$, where $\overline{n}(\br)$ is the average
number density of galaxies at position $\br$, and $P_0$ is the
power spectrum to be estimated.  For speed reasons, $P_0$ was
fixed at $5000\, h^{-3}\,$Mpc$^3$ so that a Fast Fourier transform
could be employed.  This is optimal for mimimising the variance at
$k\simeq 0.1\, h\,$Mpc$^{-1}$, whereas our signal comes from
wavenumbers with a smaller power, but in fact altering $P_0$
hardly changes the results.

\subsection{Estimating the bispectrum}

In this Section we describe how we estimate the bispectrum from
the data, taking into account the survey shape, selection
function and shot noise.

We follow \scite{FKP94} and \scite{MVH97}, and transform the field
\begin{equation}
F(\br) \equiv \lambda w(\br)\left[n(\br)-\alpha n_r(\br)\right]
\end{equation}
where $\lambda$ is a constant to be determined, $n_r(\br)$ is the
number density of a random catalogue with the same selection
function as the real catalogue, but with $1/\alpha$ times as many
particles.  If we set $\lambda = I_{22}^{-1/2}$, where
\begin{equation}
I_{ij} \equiv \int d^3\br\, w^i(\br)\, \overline n^j(\br)
\end{equation}
\cite{MVH97}, then the power spectrum may be estimated from
\begin{equation}
\langle |F_\bk|^2\rangle = P_g(\bk) + {I_{21}\over I_{22}}(1+\alpha)
\label{Power}
\end{equation}
and the bispectrum from
\begin{eqnarray}
\langle F_{\bk_1} F_{\bk_2} F_{\bk_3}\rangle\!\! &\! =\! &\!\! {I_{33}\over
I_{22}^{3/2}}\left\{B_g({\bk_1},{\bk_2},{\bk_3})+
\right.\\\nonumber & &
\!\!\!\!\!\!\!\!\!\!\!\!\!\!\!\!\!\!\!\!\!\left.{I_{32}\over
I_{33}}\left[P_g(\bk_1)+P_g(\bk_2)+P_g(\bk_3)\right]
+(1-\alpha^2){I_{31}\over I_{33}}\right\}.
\end{eqnarray}
An underlying assumption is that the power spectrum is roughly
constant over the width of the survey window function in $\bk$ space.
Because of the rather flat geometry of the survey regions, and the
holes due to star drills, this criterion is not satisfied in detail.
In practice, we have used multiple mock catalogues \citep{Coleetal98}
with the same selection criteria as the survey regions to check that
this assumption does not feed through into a biased estimate of the
bias parameters.  Also, the power spectrum and bispectrum estimates
are convolved with the window function This can lead to changes in
shape from convolution, and erroneous interpretation of correlated
noise as real features in the power spectrum.  These effects are,
however, not important in the wavenumber range ($0.1 \ls k \ls 0.5 \,
h\,$Mpc$^{-1}$) which we use for the bispectrum analysis
(\pcite{Percival2001}).

To compute the $F_\bk$, we generated random catalogues with
approximately 5 times as many particles as the real catalogue,
and analysed the North and South Galactic Pole regions (NGP \&
SGP) separately. We ignored the random fields present in the
2dFGRS as these complicate the window function and add very
little information for the current analysis. Fast Fourier transforms were performed
on a $512\times 512 \times 256$ grid which encompassed all the
particles, leading to a grid spacing of about 1 $h^{-1}\,$Mpc.

\begin{figure*}
\begin{center}
\setlength{\unitlength}{1mm}
\begin{picture}(90,55)
\includegraphics{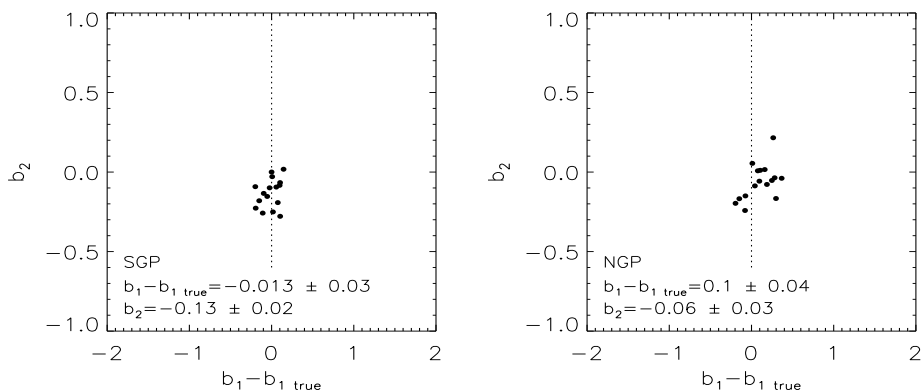}
\end{picture}
\end{center}
\caption{Error on the linear bias parameter from $\Lambda$CDM mock galaxy
catalogues in the SGP (left) and NGP (right). The average bias in
the estimator is consistent with zero  for the SGP: $-0.01 \pm
0.03$, but shows a small bias of $0.10\pm 0.04$ for the NGP. The
sample rms of $0.11$ for the SGP and $0.16$ for the NGP are used
in the analysis of the 2dFGRS.} \label{fig:Deltab}
\end{figure*}

\subsection{Choice of triangles}

We use the real parts of $F_{\bk_1} F_{\bk_2} F_{\bk_3}$ as our data,
for triangles in $\bk$ space (i.e. where $\bk_1+\bk_2+\bk_3={\bf 0}$).
Each triangle allows us to estimate a linear combination of the
parameters natural to this analysis: $c_1\equiv 1/b_1$ and $c_2\equiv
b_2/b_1^2$, through equations (6) and (7).  Note that we must use
triangles of different shape (and hence different ${\rm
Ker}[\bk_1,\bk_2]$) to lift the degeneracy between nonlinear gravity
and nonlinear bias.  As explained in the introduction, this is
equivalent to analysing the shapes of structures, which are different
in the two cases.

There is clearly a huge number of possible triangles to investigate
(many millions), and we are faced with a problem of how to choose the
triangles to analyse. The problem is that triangles which have a
wavevector in common will be correlated, through cross-terms in the
6-point function. A likelihood analysis as originally suggested in
\scite{MVH97} and \scite{VHMM98} with millions of correlated data is
unfeasible, so we take a different approach. We use two sets of
triangles of different configurations: one set with one wavevector
twice the length of another, and another set with two wavevectors of
common length.  For the former, calibrations with mock catalogues give
$\alpha_V=1.8$, and the latter $\alpha_V=1.0$.  We set a lower limit
to the wavenumber range of $k=0.1\, h\,$Mpc$^{-1}$, to avoid the
effects of convolution with the window function
\cite{Percival2001}. We set an upper limit of $k=0.5\, h\,$Mpc$^{-1}$
where, for the 2dFGRS power spectrum, shot noise begins to dominate
the signal so that little further is gained by increasing the
limit. In addition, we impose a constraint $k<0.35\, h\,$Mpc$^{-1}$
for the second configuration choice, where perturbation theory for
this configuration begins to break down. This leaves us with 80
million triangles. 
Two considerations motivate us to consider "only" these 80 million
triangles:
{\it i}) Adding more highly correlated triangles complicates the analysis
significantly and does not add much signal and 
{\it ii} ) more importantly, only these two configurations have been extensively
tested  against fully non-linear N-body simulations.
In fact not only perturbation theory may breakdown on different scales
depending on the triangle configuration, but also  the (shape dependent)
redshift space distortion parameter $\alpha_V$ (eq. 6) has been calibrated
only for these two configurations.

To cope with so many triangles, we define a new
estimator for $b_1$ and $b_2$ in the Appendix.  Although the estimator
is not minimum-variance, it is unbiased and has the big advantage that
it allows us to analyze many triangles, thus increasing
the signal to noise. The estimator does not give error bars; these are
obtained by Monte Carlo simulation from 16 mock 2dFGRS catalogues (see
Section 4). In Fig. \ref{fig:bispectrum} we show the measured
(redshift-space) 2dFGRS bispectrum from the SGP and NGP for the two
chosen triangle configurations. The dotted line shows the perturbation
theory prediction for $b_1=1$, $b_2=0$ while the dashed line shows the
shot noise contribution.  For the bispectrum, the shot noise
contribution becomes dominant around $k = 0.5\,h\,$Mpc$^{-1}$.

\section{Mock catalogues and tests}

We have used 16 mock catalogues from a Hubble Volume N-body simulation
with $\Omega_m =0.3$, $\Omega_\Lambda=0.7$ (`$\Lambda$CDM' model),
with the same selection function as the 2dF galaxy redshift survey
(\pcite{Coleetal98}) \footnote{See {\tt
http://star-www.dur.ac.uk/\mbox{$\sim$}cole/mocks/hubble.html} We find that 16 mock catalogs are sufficent to estimate the error bars.  The choice of the cosmological model is not
important for the error estimate, however this choice of the cosmological
model turns out to be not too far away from the model recovered "a
posteriori".}.  This
includes both the radial selection function and an angular mask that
reflects the varying completeness of the survey in February 2001.  The
catalogues contain mock galaxies whose positions are determined
according to the prescription described in \scite{Coleetal98}. This is
a 2-parameter exponential model based on the final density field,
i.e. it does {\it not} conform to our assumption of equation
(\ref{eq:bias}).  We will show that nevertheless the bispectrum method
recovers the bias parameter $b_1$, defined by the square root of the
ratio of the galaxy and matter power spectra\footnote{For a more
general discussion of definitions of bias, see \scite{Lahavetal2001}}.
This is the crucial test, since it is this ratio that we wish to
determine, as we can then translate the galaxy power spectrum to the
underlying mass spectrum.  For the simulated galaxies in the Hubble
volume as a whole the square root of the power spectrum ratio varies
between about 0.9 at large scales and 0.75 at $k=0.5\, h\,$Mpc$^{-1}$.
Over the scales probed by our bispectrum triangles, the bias is
roughly 0.8, but with an uncertainty which ultimately limits our error
determination.  The power spectrum of the simulated 2dFGRS catalogues
varies a little, so we estimate $b_1\equiv (P_g/P)^{1/2}$ individually
for each catalogue (and individually for SGP and NGP), from the
wavenumber range $0.1-0.5 h\,$Mpc$^{-1}$.  This sets $\beta$ for each
sample, and we fit the pairwise velocity dispersion $\sigma_p$
individually by requiring a good fit to the redshift-space power
spectrum, using the real-to-redshift mapping of equation (9).

\begin{figure}
\begin{center}
\setlength{\unitlength}{1mm}
\begin{picture}(90,55)
\includegraphics{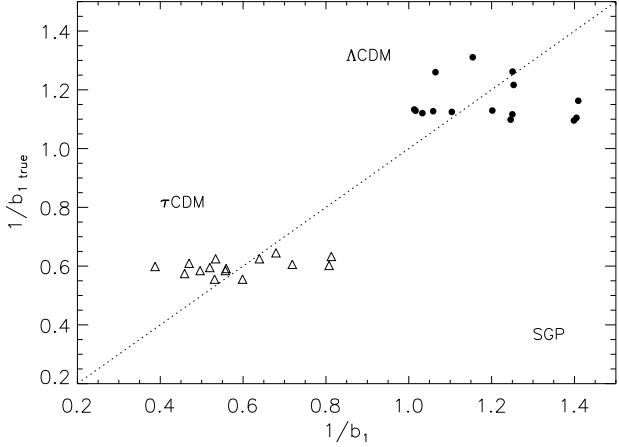}
\end{picture}
\end{center}
\caption{$1/b_1$ recovered with the bispectrum method versus the
underlying (true) $1/b_1\equiv \sqrt{P/P_g}$ for 16 mock SGP simulations
for the $\tau$CDM and $\Lambda$CDM models.  $1/b_1$ is the natural
quantity in the analysis of the bispectrum (see equation
\ref{eq:bispectrum1}). Note that the 2dFGRS has data in the NGP and
SGP, reducing the error bar compared to these mock catalogues.}
\label{fig:tCDM}
\end{figure}

We then analyse the set of triangles for each simulation, as detailed
above, and in Fig. \ref{fig:Deltab} we show the error in the bias from
16 simulations in the SGP and NGP.  The average bias in the estimator
is consistent with zero for the SGP: $-0.01 \pm 0.03$, but shows a
small bias of $0.10\pm 0.04$ for the NGP\footnote{This is
understandable since the NGP is smaller than the SGP, thus the effects
of the convolution with the window are more
important.}. Fig. \ref{fig:Deltab} shows that the approximations made
in the analysis (e.g. the functional form of the bias, the window
function, the distant observer approximation etc.) do not
significantly bias the result. We will use the sample rms of $0.11$
for the SGP and $0.16$ for the NGP as the errors in the analysis of
the 2dFGRS.  We also carried out the same analysis on 16 mock
catalogues of the SGP obtained from the Hubble volume simulation for a
$\tau$CDM model. The true underlying bias parameter is correctly
recovered with a 20\% error ($b=1.7\pm 0.3$). The error in the natural
parameter $1/b_1$ is comparable for the $\Lambda$CDM and $\tau$CDM
models, but the value of $b_1$ itself is larger in the latter case. In
Fig. \ref{fig:tCDM} we show $b_1$ recovered with the bispectrum method
versus the underlying (true) $b_1\equiv \sqrt{P_g/P}$ for 16 mock SGP
simulations for the $\tau$CDM and $\Lambda$CDM models.

\section{Results}

We analyse the catalogue as compiled in February 2001 (i.e. the
same catalogue as in \pcite{Percival2001}), and we analyse the
NGP and SGP regions separately; we excise the random fields
entirely. This leaves us with 75792 and 51862 galaxies in the SGP
and NGP respectively.  The two survey regions give us a useful
consistency check.  

We initially present results fixing $\beta=0.43$ and
$\sigma_p=385/H_0$ km$\,$s$^{-1}$ (3.85 $h^{-1}\,$ Mpc).  These are
the best-fitting values from the analysis of the 2dFGRS redshift-space
correlation function (\pcite{Peacock2001,Norberg2001a}; see also
\pcite{THX01}). Note that this value of $\beta$ involves a bias
parameter that is defined somewhat differently from ours, in terms of
the correlation function.  This bias parameter will coincide with our
$b_1$ only if certain assumptions hold, such as if the overdensity in
galaxies is everywhere $b_1$ times the mass overdensity. In due
course, $\beta$ can be determined by analysis of the 2dFGRS power
spectrum itself, but for the time being we assume that the two bias
parameters are the same.  Finally, we will marginalise over $\beta$
and $\sigma_p$.

\subsection{Bias parameters}

\begin{figure}
\begin{center}
\setlength{\unitlength}{1mm}
\begin{picture}(90,55)
\includegraphics{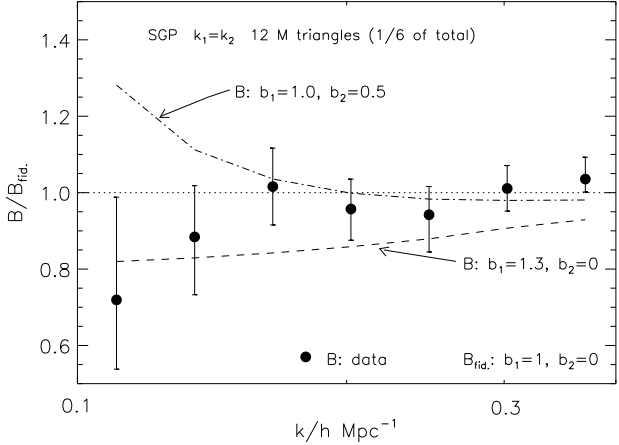}
\end{picture}
\end{center}
\caption{Ratio of the average measured bispectrum
and the average perturbation theory predictions, relative to the bispectrum
for a fiducial
unbiased model ($B_{\rm fid}$).    Dashed line $b_1=1.3, b_2=0$, dot-dashed line $b_1=1.0,
b_2=0.5$. To produce this figures only the SGP data were used
and only configurations with two wavevectors of common length.
This means that only 12 million triangles were used from the
total 80 millions.  Inclusion of the remainder excludes
both models at high confidence. The error bars are obtained
via Monte Carlo from the 16 N-body simulations, and are placed
centrally on the mean of the estimates from the mock catalogues.
This illustrates the level of bias in the estimator. The figure also shows that there is no evidence of scale dependent bias.}
\label{fig:bispectrumfrac}
\end{figure}

It is difficult to find a method to display in a single figure all of
the information from the 80 million triangles used: for this reason we
display only one configuration and for the SGP alone; adding other
configurations (or the NGP) would complicate the figure.
Fig. \ref{fig:bispectrumfrac} shows the SGP data for
configurations with two wavevectors of common length, along with the
perturbation theory predictions for different combinations of the
linear and quadratic bias parameters.  The dashed and dot-dashed lines
are the perturbation theory prediction for $b_1=1.3$, $b_2=0$ and
$b_1=1.0$, $b_2=0.5$.   The error bars come from the mock catalogues,
and the offset between the centres of the error bars and the points
shows the bias in the estimator.

Fixing the values of $\beta$ and $\sigma_p$, and including all of the
triangles in the SGP gives as a raw result $b_1=1.06$, $b_2=0.01$.
The NGP linear bias is slightly higher: $b_1=1.12$, $b_2=-0.07$.
Adjusting these results with the estimator bias obtained from the mock
catalogues (Section 4), and associating an error from the mock
catalogues gives
\begin{eqnarray}
b_1 & = & 1.07 \pm 0.11 \qquad {\rm (SGP)}\\\nonumber
b_1 & = & 1.02 \pm 0.16 \qquad {\rm (NGP)}
\end{eqnarray}
and  a combined minimum-variance weighted result of
\begin{equation}
b_1 = 1.05 \pm 0.09.
\label{eq:biasresult}
\end{equation}
Note that the estimator bias does not significantly alter the results:
ignoring it increases the estimate by 0.03, much less than the
statistical error.

Our estimate of the quadratic bias parameter from the NGP and SGP
2dFGRS is
\begin{equation}
b_2 = -0.02 \pm 0.07,
\end{equation}
with SGP and NGP individually giving $b_2 = 0.01 \pm 0.09$ and $-0.07
\pm 0.11$ respectively.

Thus we see that the 2dFGRS galaxies are perfectly consistent with
tracing the underlying mass distribution on these scales
(wavelengths $\lambda=2\pi/k$ = 13 to 63 $h^{-1}\,$Mpc) i.e. $b_1$
is consistent with unity, and $b_2$ with zero.

The results depend mildly on the values of $\beta$ and $\sigma_p$
used.  Marginalising over the distribution of these quantities
estimated from the redshift-space correlation function
\cite{Peacock2001}, the final results are virtually unchanged:
$b_1=1.04\, \pm \, 0.06$, $b_2=-0.054\, \pm\, 0.04$; adding these
errors in quadrature with those of (\ref{eq:biasresult}) gives our
final estimates:
\begin{eqnarray}
b_1 & = & 1.04 \,\pm \,0.11\\\nonumber 
b_2 & = & -0.054 \,\pm
\,0.08.
\end{eqnarray}

\subsection{Luminosity-dependent and scale-dependent bias}

The 2dFGRS exhibits luminosity-dependent clustering
\cite{Norberg2001a,Norberg2001b}, so it is reasonable to ask in what
sense the galaxies in the 2dFGRS are unbiased.  We have consistently
used all the galaxies in the 2dFGRS NGP and SGP regions, to determine
$\beta$ and the bias parameters.  With the weighting scheme employed
to measure $\beta$ and $b_1$, the weighted mean luminosity is
$1.9L_*$. From studies of the correlation function,
\scite{Norberg2001a} found a luminosity dependence in the relative
bias of 2dFGRS galaxies: $b/b_*=0.85 + 0.15 L/L_*$.  If we use this
relation to adjust our recovered bias to apply to $L_*$ galaxies, we
find they could be slightly anti-biased, but they are still
consistent with $b_1=1$ ($b_{1*}= 0.92\pm 0.11$ assuming $b_1=b$).
The 2dFGRS galaxies, as a population, trace the mass extremely well.

On scale-dependence of the bias, it is worth noting that theoretical
arguments (e.g.
\pcite{MPH98,PS2000,Seljak2000,BCFBL00,Blanton00,BerlindWeinberg01})
suggest that bias is expected to be constant on large scales (larger
than a few $h^{-1}\,$Mpc) and a scale-dependent bias is plausible on
intermediate scales, but the scale dependence is expected to be weak.
We find no evidence of scale-dependent bias (see Fig. 4). Both
scale-dependence and luminosity-dependence may be explored in detail when
the catalogue is complete.

\subsection{The matter density of the Universe}

Analysis of the redshift-space correlation function allows one to
estimate $\beta= \Omega_m^{0.6}/b_1$ \cite{Peacock2001},
and this can be combined with our determination of $b_1$ to estimate
the matter density parameter $\Omega_m$.  A strength of the result is
that it is obtained from the 2dFGRS alone, independently of all other
datasets. Following the same procedure as in the previous subsection,
we marginalise over the uncertainty in $\beta$ and $\sigma_p$, to get:
\begin{equation}
\Omega_m=0.27\pm 0.06.
\end{equation}
The small size of the error bar is worth remarking on.  For each pair
of values of $\beta$ and $\sigma_p$, we obtain an estimate of $b_1$.
The estimates of $\beta$ and $b_1$ are slightly anti-correlated, which
leads to a slightly smaller scatter in $\Omega_m$ than one would
expect from the errors on $\beta$ and $b_1$.

This result is consistent with other recent determinations by other
methods such as combining CMB with large-scale structure or supernova
measurements (e.g.  \pcite{Percival2001,Jaffe2001,Efstathiou2001}) and
from early weak lensing measurements \cite{HYG2001}.

\subsection{Comparison with previous work}

Previous work has concentrated on using cumulants, although there have
been two analyses \cite{FFFS01,SFFF2001} of the bispectrum of
infrared-selected galaxies observed with the Infrared Astronomy
Satellite (IRAS; \pcite{IRAS12,PSCz}). The IRAS samples are much
smaller than the current 2dFGRS, and are also rather shallow, so shot
noise and the radial nature of the redshift distortion is more severe
than for the 2dFGRS. Nevertheless, bias estimates have been made, and
reported with remarkably small error bars.  The most accurate quoted
values are from \scite{FFFS01} $1/b_1 = 1.2^{+0.18}_{-0.19}$ for the
PSCz survey \cite{PSCz}.  This value of the linear bias parameter is
consistent with our determinations, since IRAS galaxies have a power
spectrum which is lower than optically selected galaxies, by a factor
$\sim 1.3^2$ on the relevant scales \cite{PD94,Seaborneetal}.

Studies of the skewness, kurtosis and higher-order moments of the
optical galaxy distribution (e.g. \pcite{Gaz94,SS97,HG99,HSB00})
have also shown consistency with a linear bias of unity, but here we
are able to derive both the linear bias and the quadratic bias
simultaneously.  In addition, our present results are more
accurate and convert these general indications of a low degree of bias
into a strong constraint on theoretical models.

\section{Summary and conclusions}

We have demonstrated through analysis of the bispectrum of the 2dFGRS
that the optically-selected galaxies of the sample trace the matter
density extremely well on large scales (Fourier modes with $0.1< k< 0.5
h\,$ Mpc$^{-1}$ that approximately correspond to $30 < r < 5
\,h^{-1}\,$ Mpc). Specifically, the linear bias parameter is very close to
unity, and the quadratic (nonlinear) bias is very close to zero.  
Ironically, this is exactly the assumption which used to be made
decades ago in large-scale structure analysis, but which was
questioned when the concept of biased galaxy formation was introduced
(e.g.  \pcite{Kai84,PH85,BBKS,DR94}).  Theoretical arguments suggest
that an initial bias at formation approaches unity with time, provided
that galaxies are neither created nor destroyed \cite{Fry96}, and that
the Universe does not become curvature- or vacuum-dominated in the
meantime \cite{CMP98}.  In any case, these assumptions will clearly
fail at some level.  Note that the effective depth of the survey is
$z=0.17$ and our measurement should be interpreted as the bias and
$\Omega_m$ at this epoch (for additional discussion see
\pcite{Lahavetal2001}).  Because of a tendency for bias to approach
unity with time, we do not by this measurement rule out a significant
bias of galaxies at formation time, but the unbiased nature of the
galaxies today puts a significant constraint on theoretical models of
galaxy formation.

Currently, we find no evidence of scale-dependent bias (see
Fig. \ref{fig:bispectrumfrac}).  The size of the survey at the present
time does not allow us to place strong constraints on scale dependence
or luminosity dependence of the bias parameter, or on the nature of
biasing (e.g.  deterministic, stochastic, Eulerian, Lagrangian etc.),
but these issues will be explored with the completion of the survey.
At some level nonlinear bias must appear on small scales, from the
morphology-density relation (\pcite{Dress80,HO99}), but this is
probably on too small scales for the perturbative method of this paper
to be valid.

Our measurement of the matter density of the
Universe  $\Omega_m=0.27 \pm 0.06$ should be interpreted as $\Omega_m$ at the effective redshift of the
survey. The extrapolation at $z=0$ is model-dependent, but the changes this
correction introduces are below the quoted  error-bars for reasonnable choices
of model. Our measurement is in agreement with other independent determinations
such as cosmic microwave background together with
Hubble constant constraints
(e.g. \pcite{Jaffe2001,Efstathiou2001,Freedman2001}) which also give
comparable error bars.

It is worth re-emphasising that the analysis presented here relies
{\it only} on the 2dFGRS dataset.  It is also important that we not
only conclude that the linear bias is essentially unity, but also the
quadratic (nonlinear) bias term is constrained to be very close to
zero.  Taken together, these measurements argue powerfully that the
2dFGRS galaxies do indeed trace the mass on large scales.  In
addition to our findings, joint analysis of the 2dFGRS and CMB data
\cite{Lahavetal2001} also supports our conclusion that the 2dFGRS
galaxies are unbiased. Different methods of measuring linear
bias require different assumptions, and it is remarkable that such
different methods agree on the basic cosmological model.

\section*{Acknowledgments}
LV is supported in part by NASA grant NAG5-7154, and acknowledges the
University of Edinburgh for hospitality.  AFH thanks Rutgers and
Princeton Universities for hospitality.  We are grateful to Raul
Jimenez  and David Spergel for useful discussions.  The 2dF Galaxy Redshift Survey was
made possible through the dedicated efforts of the staff of the
Anglo-Asutralian Observatory, both in creating the 2dF instrument and
in supporting it on the telescope.

\section*{Appendix}

Here we outline the new estimator for the bias parameters.  Although
not optimal, it allows many millions of triangles to be analysed.

The likelihood of $c_1\equiv 1/b_1$ and $c_2\equiv b_2/b_1^2$ from
each triangle (labelled by $\gamma$) is degenerate, since it
constrains only a linear combination of $c_1$ and $c_2$, namely
\begin{equation}
B_\gamma = R_\gamma c_1 + S_\gamma c_2 + T_\gamma
\end{equation}
where the expressions for $R_\gamma$, $S_\gamma$, and $T_\gamma$ are
at the end of this Appendix.  The bispectrum estimate $B_{\gamma}$ has an
intrinsic statistical error $\sigma_{\gamma}$ as discussed e.g. in
\scite{MVH97}.  In computing the bias parameter, we wish to weight
the determination obtained from each triangle by the
inverse of the variance of the
corresponding bispectrum.  Since this is a cumbersome expression (see
\pcite{MVH97,VHMM98}) we approximate this by its dominant Gaussian
term,
\begin{equation}
\langle|F_{\bk_1}F_{\bk_2}F_{\bk_3}|^2\rangle\! =\!
\left(P_{g1}+P_{\rm SN}\right)\!\left(P_{g2}+P_{\rm SN}\right)\!\left(P_{g3} +
P_{\rm SN}\right)
\end{equation}
where $P_{g1} =P_g(k_1)$ etc. and $P_{\rm SN}$ is the shot noise.
Inclusion of higher-order terms leaves the results practically
unchanged for this dataset, and adds considerably to the computing
time. If we now assume that this error is Gaussian distributed we
can immediately see that the likelihood contours will be lines in
the $c_1-c_2$ plane:
\begin{equation}
\ln{\cal L}(c_1,c_2) \propto -\ln\sigma_\gamma - {(B_\gamma -
R_\gamma c_1 + S_\gamma c_2 + T_\gamma)^2\over 2 \sigma_\gamma^2}.
\end{equation}
This should give an unbiased a posteriori probability for $c_1$ and
$c_2$, if we assume uniform priors. We define a non-optimal estimator
by simply multiplying these likelihoods together, ignoring the
correlations between different triangles, forming what we term the
pseudolikelihood.  The maximum of the pseudolikelihood is still
asymptotically unbiased. The error, however, cannot be determined
internally; we estimate it by Monte Carlo methods i.e. from the
dispersion of $c_1$ and $c_2$ estimates from 16 mock catalogues. This
procedure also allows us to see if the estimator is biased or not.
Moreover, by estimating the errors via the Monte Carlo method, our
final estimates do not depend on the initial assumption of Gaussian
likelihood.

The assumption of Gaussian likelihood however, makes the maximum of the
pseudolikelihood calculable analytically, yielding the following
estimates for $c_1$ and $c_2$:
\begin{eqnarray}
\widehat{c}_1 & = & {2A_{1}A_{22}-A_2 A_{12}\over A_{12}^2-4A_{11}A_{22}}\\
\widehat{c}_2 & = & {2A_{2}A_{11}-A_1 A_{12}\over
A_{12}^2-4A_{11}A_{22}},
\end{eqnarray}
where
\begin{eqnarray}
A_{11} & \equiv & \sum_\gamma {R_\gamma^2\over 2\sigma_\gamma^2}\\
A_{22} & \equiv & \sum_\gamma {S_\gamma^2\over 2\sigma_\gamma^2}\\
A_{1} & \equiv & \sum_\gamma {(T_\gamma-B_\gamma)R_\gamma\over \sigma_\gamma^2}\\
A_{2} & \equiv & \sum_\gamma {(T_\gamma-B_\gamma)S_\gamma\over \sigma_\gamma^2}\\
A_{12} & \equiv & \sum_\gamma {R_\gamma
S_\gamma\over\sigma_\gamma^2}.
\end{eqnarray}
We can also, if desired, compute $c_1$ on the assumption that
$c_2=0$:
\begin{equation}
\widehat{c}_1 = -{A_{1}\over 2 A_{11}}\qquad (c_2\equiv 0).
\end{equation}

The expressions for $R_{\gamma}$, $S_{\gamma}$, $T_{\gamma}$ can
easily be obtained from the expression for the bispectrum of
\scite{VHMM98}; Sections 2.3 and 2.4.  We report them here for
completeness. To obtain their corresponding real space quantities
just set $\sigma_v=0$ and $\beta=0$.

\begin{eqnarray}
R_{\gamma}& = & \left\{[J(\bk_1,\bk_2)+\mu^2\beta
K(\bk_1,\bk_2)]2P_g(k_1)P_g(k_2)\right.\\ \nonumber
& \times & \left.(1+\beta\mu_1^2)(1+\beta \mu_2^2)+{\rm cyc.}\right\}D_3
\end{eqnarray}
where $\mu=-\mu_3$ for the term explicitly written and the expression for
the
kernel $K$ can be found e.g. in Catelan \& Moscardini (1994),

\begin{eqnarray}
S_{\gamma}=[(1+\beta\mu_1^2)(1+\beta\mu_2^2)P_g(k_1)P_g(k_2)+\mbox{\rm
cyc.}]D_3
\end{eqnarray}

\begin{eqnarray}
T_{\gamma}\!\!&\!\!=\!\!&\!\left\{\left[\mu_1^2\mu_2^2\beta+\frac{\beta}{2}(\mu_1^2+\mu_2^2)\right.\right.+\beta
  \mu_1 \mu_2\left(\frac{k_1}{k_2}+\frac{k_2}{k_1}\right)\\ \nonumber
  & &  \!\!\!\!\!\!\!\!\!\!\!\!\!\!+ \left.\frac{\beta^2}{2}\mu_1
\mu_2\left(\mu_1^2
     \frac{k_1}{k_2}+\mu_2^2\frac{k_2}{k_1}\right)\right]2
P_g(k_1)P_g(k_2) \\
  \nonumber
&\times &\left. (1+\beta
     \mu_1^2)(1+\beta \mu_2^2) +{\rm cyc.}\right\}D_3 \\ \nonumber
 & &\!\!\!\!\!\!\!\!\!\!\!\!\!\!+\left[P_g(k_1)\frac{(1+\beta
\mu_1^2)^2}{(1+\sigma_v^2/2k_1^2\mu_1^2)}+{\rm cyc.}\right]\frac{I_{32}}{I_{33}}+(1-\alpha^2)\frac{I_{31}}{I_{33}},
\end{eqnarray}
where $D_3$ denotes the damping term due to the incoherent small scale
velocity dispersion (see eq. 6):
\begin{eqnarray}
D_3& = &\left[ \left(1+{\alpha_V^2k_1^2\sigma_p^2\mu_1^2\over
2}\right)
\left(1+{\alpha_V^2k_2^2\sigma_p^2\mu_2^2\over 2}\right)\right.\\\nonumber
&\times &\left. \left(1+{\alpha_V^2k_3^2\sigma_p^2\mu_3^2\over 2}\right)\right]^{-1/2}.
\end{eqnarray}

\end{document}